\documentclass[preprint2,longabstract]{aastex}

\shorttitle{THE DARK MATTER HALO PROFILE OF MALIN 1}
\shortauthors{SEIGAR}

\begin{document}


\title{A cosmologically motivated description of the dark matter halo profile for the Low Surface Brightness Galaxy, Malin 1}


\author{Marc S.\ Seigar}
\affil{Department of Physics \& Astronomy, University of Arkansas at Little Rock, 2801 S.\ University Avenue, Little Rock, AR 72204}
\affil{Arkansas Center for Space and Planetary Sciences, 202 Old Museum Building, University of Arkansas, Fayetteville, AR 72701}



\begin{abstract}
In this paper we derive a possible mass profile for the low surface brightness
galaxy, Malin 1, based upon previously published space-based and ground-based
photometric properties and kinematics. We use
properties of the bulge, normal disk, outer extended disk and \ion{H}{1} mass
as inputs into mass profile models. We find that the dark matter halo model
of Malin 1 is best described by a halo profile that has undergone
adiabatic contraction, inconsistent with the findings for most disk galaxies to
date, yet consistent with rotation curve studies of M31. More
importantly, we find that Malin 1 is baryon dominated in its central regions 
out to a 
radius of $\sim10$ kpc (in the bulge region). 
Low-surface brightness galaxies are often 
referred to as being dark matter dominated at all radii. 
If this is the case, then Malin 1 would seem to 
have characteristics similar to those of
normal barred disk galaxies, as suggested by other recent
work. We also find that Malin 1 also falls on the rotation curve shear versus 
spiral arm pitch angle relation for normal galaxies, although more LSB galaxies
need to be studied to determine if this is typical.
\end{abstract}





\section{Introduction}

\begin{deluxetable}{ll}
\tabletypesize{\scriptsize}
\tablecaption{Properties of the bulge, inner and outer disk and dark matter halo of Malin 1}
\tablewidth{0pt}
\label{tab1}
\tablehead{
\colhead{Property} & \colhead{Measurement}\\
}
\startdata
Bulge effective radius$^1$, $R_e$                      & 0.6 kpc \\
Bulge effective surface brightness$^1$, $\mu_e$        & 16.8 mag arcsec$^{-2}$\\
Inner disk scalelength$^1$, $h_{\rm in}$                   & 4.8 kpc \\
Inner disk central surface brightness$^1$, $I_{\rm in_0}$  & 20.1 mag arcsec$^{-2}$\\
Outer disk scalelength$^2$, $h_{\rm out}$                  & 53 kpc\\
Outer disk central surface brightness$^2$, $I_{\rm out_0}$\hspace*{1.5cm} & 24.8 mag arcsec$^{-2}$\\
\ion{H}{1} mass$^3$, $M_{\rm HI}$                          & $(6.8\pm0.7)\times10^{10} M_{\odot}$\\
Spiral arm pitch angle, $P$                            & $25\fdg0\pm1\fdg0$\\
Halo concentration, $c_{\rm vir}$                      & 8\\
Halo virial mass, $M_{\rm vir}$                        & $2.6\times10^{12} M_{\odot}$\\
\enddata
\tablenotetext{1}{From Barth (2007) who used an HST/WFPC F814W ($I$-band) image of Malin 1 to perform a two-dimensional structural decomposition into bulge and disk components.}
\tablenotetext{2}{From Moore \& Parker (2007) who determine properties of the outer disk from a deep ground-based $R$-band image of Malin 1.}
\tablenotetext{3}{From Pickering et al.\ (1997).}
\end{deluxetable}

Malin 1 is a highly unusual disk galaxy characterized by an enormous 
\ion{H}{1} rich and extremely low surface brightness disk (Bothun et al.\ 1987;
Pickering et al.\ 1997). It has the largest radial extent of any known spiral
galaxy, with low surface brightness emission extending out to $\sim100$ kpc,
and its disk was found to have an extrapolated central surface brighness of 
only $\mu_0\simeq25.5$ mag arcsec$^{-2}$ in the $V$-band (Bothun et al.\ 1987;
Impey \& Bothun 1989), with an exponential disk scalelength of $\sim50-70$ kpc
(e.g., Moore \& Parker 2007). Although it has a very low surface brightness,
its optical luminosity is $M_v\simeq-22.9$ mag (Pickering et al.\ 1997), due to
its large extent. It also has an extremely high gas mass, with an estimated
\ion{H}{1} mass of $\sim7\times10^{10}M_{\odot}$ (Pickering et al.\ 1997; 
Matthews et al.\ 2001). As a result Malin 1 is often considered a Low Surface
Brightness (LSB) galaxy. However, recent studies of Malin 1 have started 
to highlight features more typical of normal disk galaxies (e.g., Barth
2007). The analysis of a {\em Hubble Space Telescope} (HST) WFPC2 F814W 
($I$-band) image 
presented by Barth (2007) shows that Malin 1 possesses an inner
normal stellar disk, with characteristics similar to those in regular disk 
galaxies. They calculate an exponential disk scalelength of $\sim 5$ kpc
and a disk central surface brightness of $\sim20$ mag arcsec$^{-2}$. These
data suggest that Malin 1 has characteristics similar to those of 
normal disk galaxies, in particular
barred lenticular galaxies (SBOs), which typically
show an outer disk with a larger disk scalelength (Aguirre et al.\ 2005).
Moore \& Parker (2007) have also recently presented a deep ground-based
image of Malin 1, which shows spiral structure in its inner disk, another
hint that Malin 1 may be more closely related to normal disk galaxies than
originally thought. Indeed, Malin 1 may have much in common with the
recently discovered class of objects that host extended ultraviolet (XUV)
disks, such as M83 (Thilker et al.\ 2005) and NGC 4625 (Gil de Paz et al.\
2005). All of these objects have apparently normal disks, but are surrounded
by very extended low surface brightness emission (sometimes missed entirely
in the optical) that shows up in the UV as a result of recent star formation.

In this paper we make use of a recently published \ion{H}{1} rotation curve
(Sancisi \& Fraternali 2007) to determine a possible mass profile for Malin 1.
We use the bulge/disk decomposition from Barth (2007) and the properties
of the outer disk from Moore \& Parker (2007) in our model. We also take
into account the \ion{H}{1} mass from Pickering et al.\ (1997). The mass
models produced show that Malin 1 is baryon dominated out to a radius of
$\sim$15 kpc. As LSB galaxies are typically dark matter dominated down to
small radii (e.g., de Blok \& McGaugh 1997; Kuzio de Naray et al.\ 2008)
on the surface it would appear that LSB may not be typical of LSB galaxies.
However, it should be noted that the studies presented by de Blok \& 
McGaugh (1997) and Kuzio de Naray et al.\ (2006, 2008) consisted of dwarf
LSB galaxies, which seem to be dark matter dominated beyond the inner
$\sim$1 kpc and may be baryon dominated within this radius. 
By extrapolation to much large giant LSB galaxies, such as
Malin 1, it is not implausible that these objects would also be baryon
dominated out to 10$-$15 kpc.
We also find that the rotation curve shear and spiral
arm structure of Malin 1 show that it sits nicely on the spiral pitch angle
versus shear relation for normal disk galaxies reported by Seigar et al.\
(2005, 2006). However, we also note that more LSB galaxies need to be
studied to determine if they typically fall on the same relation.

\section{Data}

Throughout this paper we use previously published data to determine 
characteristics of both the stellar, gaseous and dark matter components.
We use the {\em HST} {\tt WFPC2} F814W image described by Barth (2007) 
and the characteristics of the bulge and inner disk described therein.
We also use the deep ground-based $R$-band image from Moore \& Parker (2007)
to determine the spiral arm pitch angle of Malin 1 and we also use their
exponential scalelength of the outer disk of Malin 1 in our description of 
the baryonic mass profile.


\section{Mass modeling of Malin 1}

\subsection{The baryonic contribution}

\begin{figure*}[t]
\includegraphics{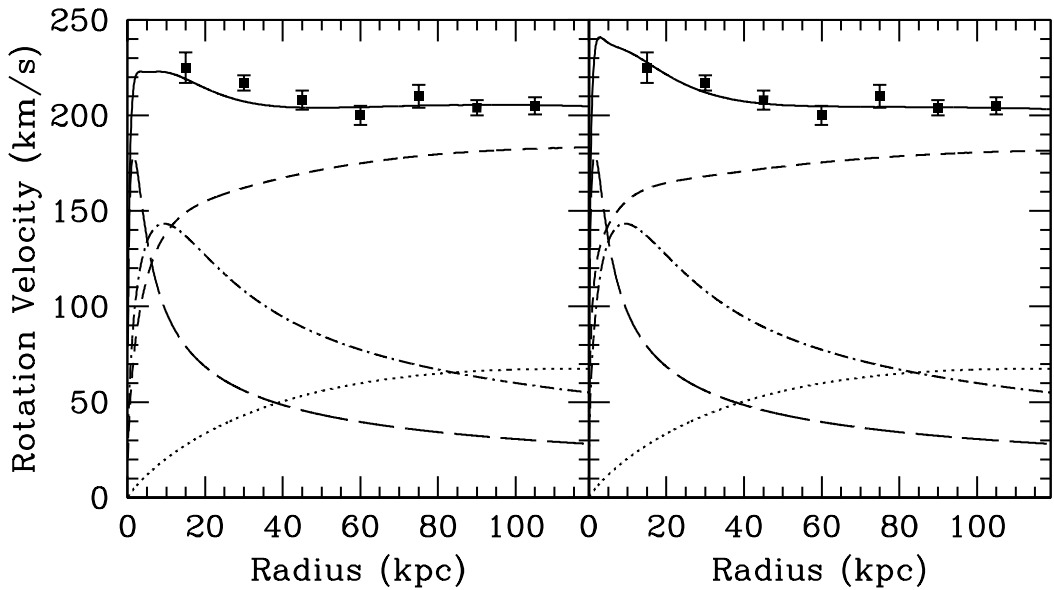}
\vspace*{8cm}
\caption{\ion{H}{1} rotation curve data from Sancisi \& Fraternali (2007) with best fitting model rotation curve (solid line) overlaid. Also plotted are the contributions from the bulge (long dashed line), the inner stellar disk (dot-dashed line), the outer \ion{H}{1}+stellar disk (dotted line) and the dark matter halo (short-dashed line). {\em Left panel}: non-AC model; {\em Right panel}: AC model.}
\label{malin1rotn}
\end{figure*}

Our goal is to determine a cosmologically motivated mass model for
Malin 1. In order to estimate the baryonic contribution to the rotation
curve, we use published bulge, inner (stellar) 
disk and outer (gas) disk properties.
We then determine several possible mass models and determine the model
that best describes the observed \ion{H}{1} rotation curve, by minimizing
the reduced-$\chi^2$.

The characterstics of the bulge and inner disk are taken from Barth (2007)
who performed a 2-dimensional bulge/disk decomposition of Malin 1, based
on an {\em HST} {\tt WFPC2} F814W ($I$-band) image. We then use the 
characteristics of the outer disk as determined from a deep ground-based 
$R$-band image presented by Moore \& Parker (2007). The characteristics
of these components are listed in Table 1.

We then assign masses to the bulge, inner disk and outer disk of Malin 1.
In order to do this we have made use of the study of 7 giant LSBs from
Sprayberry et al.\ (1995). Using their data, we determine 
typical colors for the bulge and disk components for giant LSB galaxies
and apply these same colors to the bulge and disk components of Malin 1
(assuming the inner and outer disk have similar colors). We find that a
typical bulge color for LSB galaxies is $B-R=1.5\pm0.4$ and a typical
disk color is $B-R=1.2\pm0.2$. Based upon these colors we determine
a range of calibrated stellar mass-to-light ($M/L$) ratios for the 
$I$-band and $R$-band from Bell et al.\ (2003) for the bulge, 
$(M/L_I)_{\rm bulge}$, the inner disk, $(M/L_I)_{\rm disk}$, and the 
outer disk, $(M/L_R)_{\rm disk}$. In our models we allow
mass-to-light ratios in the ranges of $1.5<(M/L_I)_{\rm bulge}<3.8$
$1.2<(M/L_I)_{\rm disk}<2.1$ 
(measured in $I$-band solar
units) and $1.3<(M/L_R)_{\rm disk}<2.7$ (measured in $R$-band solar
units), and we allow the mass-to-light ratios to vary in these ranges in
steps of 0.1. 
Large ranges in mass-to-light ratios are used in order to take into
account the large scatter in the relationships presented by Bell et al.\
(2003). We use the bulge, inner disk and outer disk light profiles to
determine the stellar mass contribution 
$M_{*}=(M/L_I)L_{\rm bulge}+(M/L_I)L_{\rm in}+(M/L_R)L_{\rm out}$, where
$L_{\rm bulge}$ is the I-band luminosity of the bulge, $L_{\rm in}$ is the
I-band luminosity of the inner disk and $L_{\rm out}$ is the R-band
luminosity of the outer extended disk.
This outer extended disk was seen in the deep imaging of Moore \& Parker 
(2007) to extend to at least a radius of 124 kpc. This disk is dominated by 
\ion{H}{1} gas and is estimated to have a mass of 
$(6.8\pm0.7)\times10^{10}M_{\odot}$ (Pickering et al.\ 1997). In this paper we 
also take into account this gas mass of the outer \ion{H}{1} disk and assume 
that it follows the same exponential disk scalelength of 53 kpc as the low 
surface brightness $R$-band disk determined by Moore \& Parker (2007). 
We also add in a stellar component (as described above) based upon the
$R$ band surface brightness measurements of Moore \& Parker (2007) and the
above $M/L$ values. It turns out that the stellar mass and the gas mass in
the outer disk are approximately equal.

\subsection{Modeling the dark matter halo}

We now explore a range of allowed dark matter halo masses
and density profiles, adopting two extreme models for disk galaxy
formation. In the first we assume that the dark matter halo
surrounding Malin 1 has not responded significantly to
the formation of a disk, i.e., adiabatic contraction (AC) does
not occur. We refer to this as our ``non-AC'' model. In this case,
the dark matter contribution to the  rotation curve is described
by a density profiles that mirrors those found in dissipationless
dark matter simulations,
\begin{equation}
\rho(r)=\frac{\rho_{s}}{(r/r_s)(1+r/r_s)^2},
\label{NFW}
\end{equation}
where $r_s$ is a characteristic ``inner'' radius, and $\rho_s$ is a 
corresponding inner density. Here we have adopted the profile shape
of Navarro et al.\ (1996; hereafter NFW). The NFW profile is a
two-parameter function and is completely specified by choosing
two independent parameters, e.g., the virial mass $M_{\rm vir}$ (or virial
radius $R_{\rm vir}$) and concentration $c_{\rm vir}=R_{\rm vir}/r_s$ define 
the profile
completely (see Bullock et al.\ 2001b for a discussion). Similarly,
given a virial mass $M_{\rm vir}$ and the dark matter circular velocity at
any radius, the halo concentration $c_{\rm vir}$ is completely determined.

In the second class of models we adopt the scenario of adiabatic
contraction (AC) discussed by Blumenthal et al.\ (1986; see
also Bullock et al.\ 2001a and Pizagno et al.\ 2005). Here we assume
that the baryons and dark matter initially follow an NFW
profile and that the baryons cool and settle into the halo center
slowly compared to a typical orbital time. This slow infall provokes
an adiabatic contraction in the halo density distribution
and gives rise to a more concentrated dark matter profile. The
idea of adiabatic contraction was originally discussed as to explain
the ``conspiracy'' between dark halos and disk sizes that
gives rise to a featureless rotation curve (Rubin et al.\ 1985) but
has since proven to be remarkably accurate in describing the formation
of disk galaxies in numerical simulations (e.g., Gnedin
et al.\ 2004, and references therein), although the degree to which
this process operates in the real universe is currently uncertain.
For example, Dutton et al.\ (2005) showed that adiabatic contraction
models are inconsistent with the rotation curves measured
and the expected NFW concentrations for a sample of six galaxies.
They suggest that mechanisms such as stellar feedback and
stellar bars may result in less concentrated halos than predicted
by adiabatic concentration.

In our AC model we take the contraction into account following
the prescription of Blumenthal et al.\ (1986). Note that Gnedin
et al.\ (2004) advocate a slightly modified prescription, but the
differences between the two methods are small compared to the
differences between our AC model and our non-AC model. In principle,
any observational probe that can distinguish between AC and
non-AC-type scenarios provides an important constraint on
the nature of gas infall into galaxies (i.e., was it fast or was it slow?).

We iterate over the central and $\pm1\sigma$  values
found in the bulge-disk decompositions for $h$ and $L_{\rm disk}$ and explore
the values of mass-to-light ratio discussed above,
for the bulge $1.5<(M/L_I)_{\rm bulge}<3.8$ for the inner
disk  $1.2<(M/L_I)_{\rm disk}<2.1$ and for the outer disk 
$1.3<(M/L_R)_{\rm disk}2.7$. In each case we assume average
values for
$(M/L_I)_{\rm bulge}$, $(M/L_I)_{\rm disk}$ and $(M/L_R)_{\rm disk}$. 
For each choice of bulge-inner disk-outer disk 
model parameters and mass-to-light ratios, we allow the (initial) halo
NFW concentration parameter to vary over the range of viable
values, $c_{\rm vir}= 3 - 31$ (Bullock et al.\ 2001b). We then determine
the halo virial mass $M_{\rm vir}$ necessary to reproduce the rotation velocity
at 2.2 inner disk scalelengths ($V_{2.2_{\rm in}}=10.56$ kpc) and the rotation 
velocity at 2.2 outer disk scalelengths ($V_{2.2_{\rm out}}=116.6$ kpc)
for the galaxy and determine the implied fraction of the mass in
the system in the form of stars compared to that ``expected'' from
the Universal baryon fraction, $f_{*}=M_{*}/(f_{b}M_{\rm vir})$. We make the
(rather loose) demand that $f_{*}$ lies within the range of plausible
values $0.01f_{b}<f_{*}<f_{b}$.

\begin{deluxetable}{lcc}
\tabletypesize{\scriptsize}
\tablecaption{Malin 1 best fitting models.} 
\tablewidth{0pt}
\label{tab2}
\tablehead{
\colhead{Parameter} & \colhead{non-AC} & \colhead{AC}\\
}
\startdata
Shear                                    & 0.50$\pm$0.01      & 0.47$\pm$0.01      \\
NFW concentration, $c_{\rm vir}$         & 15                 & 8                  \\
Virial mass, $M_{\rm vir}$ ($M_{\odot}$) & $1.8\times10^{12}$ & $2.6\times10^{12}$ \\
Bulge mass-to-light ratio, $(M/L_I)_{\rm bulge}$ & 2.2                & 2.2                \\
Inner disk mass-to-light ratio, $(M/L_I)_{\rm disk}$ & 1.2                & 1.2                \\
Outer disk mass-to-light ratio, $(M/L_R)_{\rm disk}$ & 1.3                & 1.3                \\
$\chi^2/\nu$                             & 2.45               & 1.30               \\
\enddata
\tablecomments{``non-AC'' is the best-fit model to the rotation curve from Sancisi \& Fraternali (2007) without adiabatic contration. ``AC'' is the best-fit model to the same rotation curve data using the Adiabatic Contraction prescription from Blumenthal et al.\ (1986).}
\end{deluxetable}

For each chosen value of $c_{\rm vir}$ and adopted disk formation scenario
(AC or non-AC), the chosen values of $V_{2.2_{\rm in}}$ and $V_{2.2_{\rm out}}$
constraints define the rotation
curve completely and thus provide an implied shear rate
at every radius. Figure 1 shows the \ion{H}{1} rotation velocity data from
Sancisi \& Fraternali (2007) overlaid with 
best-fitting model rotation curves that
we derive for Malin 1 for both the non-AC ({\em left panel}) and AC ({\em right
panel}) models. The best fit overall rotation curve model is divided 
into its bulge, inner disk, outer disk and halo components. We find that the
best fitting rotation curve model is more consistent with a halo that has 
undergone adiabatic contraction, rather than a pure NFW model. Our 
preference for the AC model 
is inconsistent with the findings that an adiatically contracted
halo model rarely describes the observed rotation curves of disk galaxies
(e.g., Kassin et al.\ 2006a, b), yet consistent with the rotation curve of
M31, which also appears to require adiabatic contraction (Klypin et al.\ 2002;
Seigar et al.\ 2008a). 
Given the accumulation of evidence that the rotation curves of
disk galaxies (especially late-type disk galaxies with little or no bulge)
tend to be inconsistent with the predictions of AC, it is surprising that
our AC model seems to work best for Malin 1. Considering that the rotation
curve of M31 (the nearest and best-studied of galaxies) is also consistent
with the expectations of AC (e.g., Seigar et al. 2008a), maybe this suggests
that Malin 1 has properties similar to those of normal surface brightness,
bulge-dominated galaxies, of which to-date only a handful have been studied in
this manner. From here on, we adopt
our AC model as the fiducial model.
The virial mass, $M_{\rm vir}$, and concentration, $c_{\rm vir}$,
for the best fitting halo model are listed in Table 1. 
Table 2 lists parameters of both the non-AC and AC models for comparison.
Figure 2 shows 
the enclosed mass as a funtion of radius for our best-fitting AC
model, separated into bulge, inner disk, outer disk and halo components.

\begin{figure}[t]
\plotone{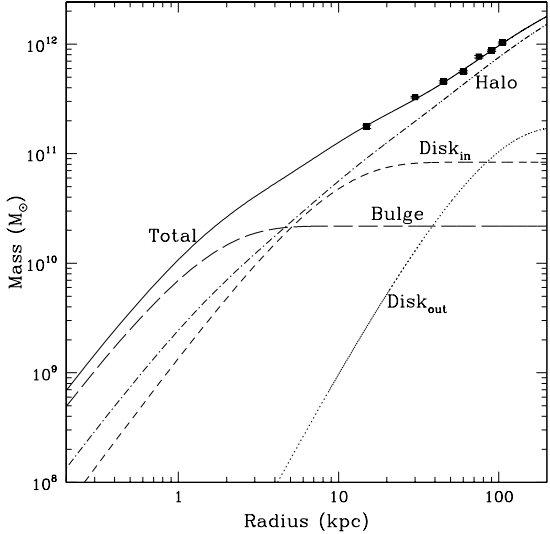}
\caption{
Total enclosed mass (solid line) as a function of radius for Malin 1 for
the best-fitting AC model. The 
enclosed mass is divided into its bulge (short-dashed line), inner disk 
(long-dashed line), outer disk (dotted line) and dark matter halo 
(dot-dashed line) components. The data points correspond to the \ion{H}{1}
rotation curve data from Sancisi \& Fraternali (2007) shown in the left 
panel of Figure \ref{malin1rotn}.}
\label{malin1mass}
\end{figure}

The most interesting aspect of the best-fitting rotation curve, is the 
fact that it appears to be dominated by the bulge in the
inner regions of Malin 1, out to a radius of $\sim7$ kpc.
As LSB galaxies are often referred to as being dark matter dominated
at all radii, on the surface this result would suggest that Malin 1 may
not be a typical LSB galaxy. However, the studies of dwarf LSB galaxies
by de Blok \& McGaugh (1997) and Kuzio de Naray et al.\ (2006, 2008)
show that these galaxies may actually be baryon dominated in their
very central $\sim$1 kpc. By extrapolation to giant LSB galaxies it 
may seem plausible that these larger counterparts may also be 
baryon dominated out to $\sim5-10$ kpc. However, a common criterion
for classifying LSB galaxies is a disk central surface brightness
fainter than $\mu_B=23.0$ mag arcsec$^{-2}$ (Impey \& Bothun 1997).
A galaxy with a disk central surface brightness fainter than this
would present a $>4\sigma$ deviation from the distribution of 
disk surface brightnesses found by Freeman (1970). As a result any
galaxy with a disk central surface brightness les than 
$\mu_B=23.0$ mag arcsec$^{-2}$
is typically classified as an LSB. However, Barth (2007) determined
a $B$ band disk central surface brightness of $\mu_B(0)=22.3$ mag
arcsec$^{-2}$ for Malin 1. This would not classify Malin 1 as an LSB
galaxy, but as an intermediate surface brightness disk, if we were to
use the classification system of McGaugh (1996). Taken together with
our mass profile, which seems to suggest that Malin 1 is baryon 
dominated out to large radii, this may be revealing that Malin 1 is
not as atypical as originally thought. It seems that Malin 1 has
characteristics that are similar to those of SBO type galaxies, but it
is also embedded in a very extended, optically faint, gas-rich outer
disk beyond its normal inner disk.

Although the inner most point of the rotation curve is at 15 kpc, it would
be difficult to model Malin 1 with any cosmologically motivated dark matter
profile that would not be baryon dominated within $\sim5$ kpc. Even a
pseudo-isothermal profile (see e.g., Simon et al.\ 2005; Kuzio de Naray et al.\
2006 for a description of the pseudo-isothermal profile) would be baryon
dominated out to a similarly large radius, as such a profile tends to provide
comparitively less dark matter at small radii. 

Given the lack of points
within 15 kpc, it is almost impossible to determine whether a NFW model
or a pseudo-isothermal model provides the best possible profile for the dark
matter halo of Malin 1. The difference between these two types of dark matter
halo profile are most sensitive in the very inner regions, where the NFW-type
profile provides a ``cuspy'' inner density profile and the pseudo-isothermal
profile provides a constant density core (see e.g., Simon et al.\ 2005).
To determine which of these best describes the halo of Malin 1, we would need
better sampled kinematics within the inner 15 kpc. Since more and more
evidence seems to suggest that pseudo-isothermal models work better for
describing the dark matter distribution in disk galaxies (e.g., Gentile et al.\
2004, 2005; Shankar et al.\ 2006; Spano et al.\ 2008) 
it seems important better sampled
spectrosocopy be observed for the inner regions of Malin 1 in any future 
study.

Of course, the use of optical data to model the stellar parts of Malin 1 is
limited. The expected stellar $M/L$
ratio in the optical has a very large scatter
(e.g., Bell \& de Jong 2001; Bell et al.\ 2003), and ideally we would prefer
to have near-infrared images of Malin 1, which would provide a more accurate
stellar $M/L$ ratio.

\section{Does Malin 1 lie on the Spiral Arm Pitch Angle versus Shear relation?}

\subsection{Measurement of the pitch angle of Malin 1}

\begin{figure}[t]
\plotone{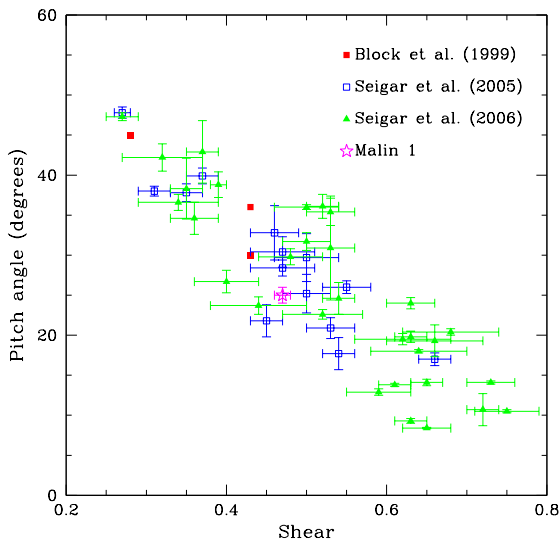}
\caption{The Pitch angle versus Shear relation from Seigar et al.\ (2005, 2006)
with Malin 1 overlaid. The red points represent data from Block et al.\ (1999);
the blue points represent data from Seigar et al.\ (2005); the green points
represent data from Seigar et al.\ (2006) and the magenta point represents
Malin 1.}
\label{malin1shear}
\end{figure}

Spiral arm pitch angles are measured using the same technique
employed by Seigar et al.\ (2004, 2005, 2006, 2008b). A two-dimensional fast-
Fourier transform technique (FFT) is used, which employs a
program described by Schr\"oder et al. (1994). Logarithmic spirals
are assumed in the decomposition. 
The amplitude of each Fourier component is given by
\begin{equation}
A(m,p)=\frac{\Sigma_{i=1}^{I}\Sigma_{j=1}^{J}I_{ij}(\ln{r},\theta)\exp{[-i(m\theta+p\ln{r})]}}{\Sigma_{i=1}^{I}\Sigma_{j=1}^{J}I_{ij}(\ln{r},\theta)},
\label{pitch}
\end{equation}
where $r$ and $\theta$ are polar coordinates, $I(\ln{r},\theta)$ is the 
intensity at position $(\ln{r},\theta)$, $m$ represents the number of arms or 
modes, and $p$ is the variable associated with the pitch angle $P$, defined
by $\tan{P}=-(m/p)$. We measure the pitch angle $P$ of the $m=2$ component. 
The resulting pitch angle measured using equation \ref{pitch} is in radians, 
and this is later converted to degrees for ease of perception.

The range of radii over which the FFT was applied
was selected to exclude the bulge (where there is no information
about the arms) and to extend out to the outer limits of
the arms in the deep R-band image of Malin 1 from Moore \& Parker (2007). The 
radial extent of
the bar was measured manually (see, e.g., Grosbol et al.\ 2004),
and the inner radial limit applied to the FFT was chosen to be outside
this radius. The physical distance was calculated using a Hubble
constant $H_{0}=73$ km s$^{-1}$ Mpc$^{-1}$ (Spergel et al.\ 2007) and the 
recessional velocity, $V_{rec}=24750\pm10$ km s$^{-1}$ (de Vaucouleurs et al.\ 
1991; hereafter RC3). 
The pitch angle was then determined from peaks in the Fourier spectrum, 
as this is the most powerful method for finding periodicity in a distribution 
(Consid\`ere \& Athanassoula 1988; Garcia-Gomez \& Athanassoula 1993). The 
radial range over which the Fourier analysis was performed was chosen by eye 
and is probably the dominant source of error in the calculation of the pitch 
angle, as spiral arms are only approximately logarithmic and sometimes abrupt 
changes can be seen in spiral arm pitch angles (e.g., Seigar \& James 1998). 

The image was first 
deprojected to face-on. Mean uncertainties of position angle 
and inclination as a function of inclination were discussed by Consid\`ere \&
Athanassoula (1988). For a galaxy with high inclination, there are clearly 
greater uncertainties in assigning both a position angle and an accurate 
inclination. These uncertainties are discussed by Block et al.\ (1999) and 
Seigar et al.\ (2005), who take a galaxy with low inclination ($<30^{\circ}$) 
and one with high inclination ($>60^{\circ}$) and varied the inclination angle 
used in the correction to face-on. They found that for the galaxy with
low inclination, the measured pitch angle remained the same.
However, the measured pitch angle for the galaxy with high inclination
varied by 10\%. Since inclination corrections are likely to be largest for 
galaxies with the highest inclinations, cases in which inclination is 
$>60^{\circ}$ are taken as the worst case scenario. Since the inclination of 
Malin 1, $i\simeq23^{\circ}$, the error in deprojecting to a face-on 
orientation is likely to be very low.

From the $R$ band image of Malin 1 presented in Moore \& Parker (2007), the
pitch angle of their overlaid spiral is measured as $P=25\fdg0\pm1\fdg0$.

\subsection{Measurement of rotation curve shear}

We use our best fit model rotation curve to the \ion{H}{1} rotation velocities 
from Sancisi \& Fraternali (2007) to measure the shear for 
Malin 1. The shear is measured using the same method used by other authors
(e.g., Block et al.\ 1999; Seigar et al.\ 2004, 2005, 2006; Seigar 2005).

Rotation curve shear is defined as,
\begin{equation}
S=\frac{A}{\omega}=\frac{1}{2}\left(1-\frac{R}{V}\frac{dV}{dR}\right),
\label{shear}
\end{equation}
where $A$ is the first Oort constance, $\omega$ is the angular velocity, and
$V$ is the rotation velocity at a radius $R$. The shear depends on the shape
of the rotation curve. For a rotation curve that remains flat, $S=0.5$, for
a falling rotation curve, $S>0.5$, and for a continually rising rotation curve,
$S<0.5$. 

Using equation \ref{shear} and the model rotation curve, 
we have calculated the shear for Malin 1 at a
radius of 10 kpc (the same radius at which Seigar et al.\ 2005, 2006 
measured their values for rotation curve shear). The dominant source of error 
on the measurement of shear is the rms error in the
rotation curve. This is typically $<10$\%. In order to calculate the
shear, the value of $dV/dR$, measured in km s$^{-1}$ arcsec$^{-1}$,
is calculated as a function of radius for the outer part
of the rotation curve (i.e., past the radius of turnover and the bulge 
component). 

Using this technique we find a rotation curve shear of Malin 1, 
$S=0.47\pm0.01$, indicating that the rotation curve for Malin 1 is declining
at this radius.

\subsection{The shear versus pitch angle relation}

From the spiral arm detected by Moore \& Parker (2007), the pitch angle of 
Malin 1 is $P=25\fdg0\pm1\fdg0$. We also find a shear of $S=0.47\pm0.01$
from the \ion{H}{1} rotation curve presented by Sancisi \& Fraternali (2007).
Figure 3 shows the result of plotting the pitch angle and shear of Malin 1
on the spiral arm pitch angle versus rotation curve shear relation from 
Seigar et al.\ (2005, 2006). As can be seen Malin 1 fits nicely on this
relation, which was originally 
determined for normal spiral galaxies. This is the
first LSB galaxy which has been plotted on the shear versus pitch angle
relation. It now seems appropriate that these measurements be made for
more LSB galaxies to see if their shear and pitch angle remain consistent
with the relation for normal brightness galaxies.

\section{Conclusions}

We conclude that Malin 1 is not as atypical as originally thought. We 
highlight the fact that its $B$ band disk central surface brightness
of $\mu_B(0)=22.3$ mag arcsec$^{-2}$ as determined by Barth (2007) 
seems to place it in the category of intermediate brightness galaxies
(McGaugh 1997). Taken together with our result here, that Malin 1
appears to be baryon dominated to $\sim$10 kpc, this may suggest that
Malin 1 has characteristics typical of normal galaxies. However, it still 
remains a very unusual galaxy, as it 
is also embedded in a very extended, gas-rich, outer disk. While
Barth (2007) compared Malin 1 to SBO galaxies, the discovery of spiral
structure in its disk (Moore \& Parker 2007) would suggest that 
Malin 1 may very well be of later-type than this. The break in the
outer disk of Malin 1 to that of a disk with a larger scalelength is
not unusual for disk galaxies (e.g., Pohlen et al.\ 2002; Erwin et 
al.\ 2005, 2007). Malin 1 may just exhibit an extreme case of this
phenomenon.

The spiral
structure and rotation curve shear of Malin 1 are both consistent with
those of normal disk galaxies, and they both fall nicely on the rotation
curve shear versus spiral arm pitch angle relation reported by Seigar et al.\
(2005, 2006). It is possible that a comparison of shear values and pitch 
angles for LSB galaxies reveal that they follow the same relation as normal
galaxies. For this reason, in the future, we intend to make these measurement 
for a large sample of LSB galaxies.

\acknowledgments
This research has made use of the NASA/IPAC Extragalactic Database (NED) 
which is operated by the Jet Propulsion Laboratory, California Institute
of Technology, under contract with the National Aeronautics and Space
Administration. The research presented in this paper has been made possible
by the Arkansas Space Grant Consortium. MSS also acknowledges the
anonymous referee, whose input greatly improved the content of this
article.

\end{document}